# The problem of positivity in 1+1 dimensions and Krein spaces

V. Montalbano[(*)]

[(*)] *Dipartimento di Fisica - Centro I.M.O. Sezione di Siena - Università di Siena - Siena - Italia*

*Riassunto*

Il problema della positività della metrica indotta nello spazio degli stati del campo scalare privo di massa in 1+1 dimensioni è stato esaminato. Si considerano due rappresentazioni negli spazi di Krein per il campo scalare libero privo di massa. Si dimostra che una rappresentazione è un caso particolare dell'altra. Vengono inoltre discussi aspetti particolari e vantaggi di entrambe le rappresentazioni.

*Summary*

The possibility of introducing a positive metric on the states of the massless scalar field in 1+1 dimensions by mean of Krein spaces is examined. Two different realisations in Krein spaces for the massless scalar field are compared. It is proved that one is a particular case of the other. The peculiarities and advantages of both realisations are discussed.

# Introduction

In order to investigate some peculiar, and usually non-perturbative, aspect of quantum field theory, bidimensional models are studied. These models have the advantage that one or more non trivial behaviour of realistic theories in 3+1 dimensions are faithfully enough reproduced without showing their full complexity.

Furthermore, a collection of models in 1+1 dimensions are exactly soluble. For example, the Schwinger model [1] (i. e. the quantum electrodynamics in 1+1 dimensions) is a soluble [2-5] gauge theory reproducing some fundamental structures of non abelian gauge theory (e. g. the axial current anomaly and the $\theta$ vacua [4,5] ).

The problem of positivity arises when the solutions of these models necessitate the use of fields represented on a space with indefinite metric. With such a metric one has problems with the reconstruction theorem. Thus, in the axiomatic approach [6] to the theory, even if one knows all the Wightman functions, it is not possible to obtain all states. In order to find all possible states, the representation space must be equipped with a positive metric. It is important in doing so that, however the positive metric is achieved, the algebra of the fields must not be changed. In particular, one can not arbitrarily restrict the set of possible test function.

It has been suggested [7-9] that a solution of this problem is the use of the Krein metric [10].

In this note the free massless scalar field in 1+1 dimensions is considered. It is well known that this field is affected by the problem of indefiniteness of the metric [6,11-13] induced by its Wightman functions.

In the following section, the indefinite metric for the massless scalar field is shown. Two different realisation in the Krein spaces [8,9,14] for this field are described: a former [8,9] realisation for the space of states and a latter one [14] that appears more general but very different from the previous representation.

In the next section, it is given the proof that the two realisation are equivalent and describe the same space of states, i. e. the former one is a particular case of the latter representation. They look different because of the former one is not given in the canonical decomposition of the Krein space.

Finally, the peculiarities and advantages of each representation are discussed.

## Massless scalar field and Krein spaces

The algebra of the free massless scalar field $\phi(f)$ with $f \in \Sigma(\mathfrak{R}^2)$ satisfies

$$\Box \phi(x) = 0, \tag{1}$$

and

$$[\phi(x), \phi(0)] = -iD^{(+)}(x), \tag{2}$$

with

$$D(x) = \frac{1}{2} \varepsilon(x_0) \theta(x^2). \tag{3}$$

The Fock representation of this algebra, given by the vacuum state

$$\langle 0|\phi(x)|0\rangle = 0, \tag{4}$$

and the two-point Wightman function

$$W(x) = \langle 0|\phi(x)\phi(0)|0\rangle = -iD^{(+)}(x), \tag{5}$$

where

$$D^{(+)}(x) = \frac{1}{4\pi i} \log(-x^2 + i\varepsilon x_0), \tag{6}$$

displays an indefinite inner product induced on the space of test functions $\Sigma(\mathfrak{R}^2)$

$$\langle f, g \rangle = \int_{\mathfrak{R}} dx^2\, dy^2 f^*(x) g(y) W(x-y) = \tag{7}$$

$$= \frac{1}{4\pi} \int_{\mathfrak{R}} \frac{dp}{|p|} [f^*(|p|,p)g(|p|,p) - f^*(0,0)g(0,0)\,\theta(1-|p|)] \tag{8}$$

(the last expression is obtained by applying the Fourier transform in eq. 8).

Let me describe the first Krein realisation proposed [8,9] for the massless scalar field. From the indefinite inner product (7) an Hilbert scalar product is given as follows

$$(f,g)_K = \langle f_0, g_0 \rangle + \langle f, \chi^* \rangle \langle \chi^*, g \rangle + f^*(0)g(0), \tag{9}$$

where the space of test functions $\Sigma(\Re^2)$ has been decomposed

$$\Sigma = \Sigma_0 + V. \tag{10}$$

The space $\Sigma_0$ is formed by the test function whose the Fourier transform is null for $p = 0$ and $V$ is a one-dimensional space generated by a real symmetric test function that satisfies the following conditions

$$\chi^*(0,0) = 1, \tag{11}$$

$$\langle \chi^*, \chi^* \rangle = 0. \tag{12}$$

The completion of the space of test functions in the Krein topology induced by the scalar product (9) is called $K$ and has been carefully studied by these authors and it is composed as follows

$$K = L^2\left(\frac{dp}{|p|}, C_+\right) \oplus V_0 \oplus V, \tag{13}$$

where $V_0 \equiv \{\lambda v_0, \lambda \in C\}$ and $v_0$ is an element of $K$ with the properties

$$(v_0, f)_K = \langle \chi^*, f \rangle, \tag{14}$$

$$\langle v_0, f \rangle = f(0), \tag{15}$$

for every test function.

In a long and carefully study in ref. [8] and [9], these authors have proved that the completion $K$ is a Krein space, i. e. it is the maximal space associated to the correlation functions of the massless scalar field. Moreover, some aspects of symmetries of this model are been pointed out by mean of computations by using the Krein metric.

More recently, another Krein metric has been proposed [14] for this field. The starting point is the definition of Krein space [10] which is recalled hereafter.

A vector space $\Im$ with a sesquilinear Hermitian form $Q$ defined on it is said to have a $Q$-metric

$$Q(f, g) = \langle f, g \rangle \qquad f, g \in \Im, \qquad (16)$$

in general indefinite.

If this space furthermore admits a canonical decomposition

$$\Im = \Im^{(+)} [+] \Im^{(-)}, \qquad (17)$$

where $\Im^{(+)}$ is orthogonal in the $Q$-metric to $\Im^{(-)}$ and both $\Im^{(+)}$ and $\Im^{(-)}$ are Hilbert spaces with norm

$$\|f\|^2 \equiv \langle f, f \rangle \qquad f \in \Im^{(+)}, \qquad (18)$$

$$\|f\|^2 \equiv -\langle f, f \rangle \qquad f \in \Im^{(-)}, \qquad (19)$$

then $\Im$ is a Krein space.

The Krein metric is introduced by demonstrating that a canonical decomposition can be defined as follows: $f = f_+ + f_-$ where

$$f_+ = f + \langle \chi, f \rangle \chi, \qquad (20)$$

$$f_- = -\langle \chi, f \rangle \chi, \qquad (21)$$

and $\chi$ is an element of $\Im$ such as

$$\langle \chi, \chi \rangle = -1. \qquad (22)$$

The inner scalar product is provided by using eq. (20)

$$(f, g) \equiv \langle f_+, g_+ \rangle + \langle f, \chi \rangle \langle \chi, g \rangle. \qquad (23)$$

Moreover, an alternative but useful form is given

$$(f, g) = \langle f, g \rangle + 2 \langle f, \chi \rangle \langle \chi, g \rangle. \qquad (24)$$

The two Krein metric proposed, eq. (9) and eq. (23) or (24), look similar but are very different. In fact, in ref. [14] the authors suggest that metric (8) is not a Krein metric.

Let me conclude this section by pointing out that this latter representation of the Krein space associated to the massless scalar field has been achieved by using few general arguments and in general computation in this case are very little involved.

## Results and discussion

The first argument that arises by comparing the two proposed metric is that the proof given in ref. [14] states that the scalar product (9) is not given in a canonical decomposition and this is not enough for excluding the Krein structure of the topology proposed.

In fact, it is possible to find an element in $K$ such that the eq. (22) is satisfied and by using the scalar product (24) the Krein metric induced has the same form of eq. (9).

Let me consider the element $\chi \in K$ defined as follows

$$\chi = \frac{1}{\sqrt{2}}(v_0 - \chi^*), \tag{25}$$

it is easy to verify that condition (21) is satisfied

$$\langle \chi, \chi \rangle = \frac{1}{2}(\langle v_0, v_0 \rangle + \langle \chi^*, \chi^* \rangle - \langle \chi^*, v_0 \rangle - \langle v_0, \chi^* \rangle) = -1, \tag{26}$$

where eq. (11) and (15) have been utilised. Moreover, it has been useful to know that (see ref. [9]) it exists an operator $\eta$ such that these relations are satisfied

$$\eta \chi^* = v_0 \qquad \qquad \eta v_0 = \chi^*, \tag{27}$$

thus $\eta^2 = 1$ and it is straightforward to show that

$$\langle v_0, v_0 \rangle = \langle \eta v_0, \eta v_0 \rangle = \langle \chi^*, \chi^* \rangle = 0. \tag{28}$$

If the element (25) is utilised in order to determine the form of the Krein scalar product in the form (24), the following expression is obtained

$$(f, g) = \langle f, g \rangle + \langle f, v_0 - \chi^* \rangle \langle v_0 - \chi^*, g \rangle =$$
$$= \langle f, g \rangle + [f^*(0) - \langle f, \chi^* \rangle][g(0) - \langle \chi^*, g \rangle] = (f, g)_K. \tag{29}$$

Thus, the result is that the Krein structure (9) proposed in ref. [8] and [9] is nothing else that a particular case of the general expression (23) given in ref. [14]. Then, the Krein space induced is the same, i. e. $\Im \equiv K$.

Let me conclude by focusing on a peculiar aspect of the realisation (9) of the Krein space. Since $v_0$ belongs to the closure of the Krein space almost every interesting calculation implies the use of series and limit. This fact explains the complex study necessary to comprehend the structure of the Krein space.

On the contrary, the general Krein realisation (24) allows very straightforward demonstrations and, if one chooses a non boundary element for implementing it, computations are easily done. For example, in ref. [14] it is demonstrated that in this model the interpretation for the Fourier components as probability amplitudes for the momentum operator is lost.

Thus, the realisation (9) is indicated when one wants to study the completeness of the space or the boundary properties. In all other case, the realisation (24) is more advantageous.